\documentclass[twocolumn,showpacs,amsmath,amssymb,prb]{revtex4} 
\newcommand{\be}{\begin{equation}}
\newcommand{\ee}{\end{equation}}
\newcommand{\bea}{\begin{eqnarray}}
\newcommand{\eea}{\end{eqnarray}}

\def\nn{\nonumber\\}
\def\r#1{({\ref{#1}})}

\usepackage{graphicx}
\usepackage{epsf}
\usepackage{psfrag}

\begin{document}

\title{Dynamical structure factor of the anisotropic Heisenberg chain
  in a transverse field}

\author{Jean-S\'ebastien Caux$^1$, Fabian H.L. Essler$^2$ and Ute L\"ow$^3$}

\affiliation{
$^1$ ITFA, U. of Amsterdam,
Valckenierstraat 65, 1018 XE Amsterdam, The Netherlands \\
$^2$ Department of Physics, Brookhaven National Laboratory, Upton, NY
11973-5000, USA\\
$^3$Institut f\"ur Theoretische Physik, Universit\"at zu K\"oln,\\
Z\"ulpicher Stra\ss e 77, 50937 K\"oln, Germany}

\date{\today}

\begin{abstract}
We consider the anisotropic Heisenberg spin-$1/2$ chain in a
transverse magnetic field at zero temperature. We first determine all
components of the dynamical structure factor by combining exact
results with a mean-field approximation recently proposed by Dmitriev
{\it et al}., JETP 95, 538 (2002). We then turn to the small
anisotropy limit, in which we use field theory methods to obtain exact
results. We discuss the relevance of our results to Neutron scattering
experiments on the 1D Heisenberg chain compound ${\rm Cs_2CoCl_4}$.
\end{abstract}


\maketitle
\section{Introduction}
The spin-$\frac{1}{2}$ Heisenberg chain is one of the most
well-understood paradigms for quantum critical behavior. The latter
term is coined to describe a situation in which quantum fluctuations
induce critical behavior as a parameter like doping or a magnetic
field is varied. There are many physical realisations of models of
(weakly coupled) Heisenberg chains in anisotropic
antiferromagnets. Whereas these materials usually display magnetic
long-range order at very low temperatures $T<T_N$, which is induced by
the weak coupling between chains, there is a large ``window'' above
$T_N$ where quantum fluctuations dominate and scaling associated with
a quantum critical disordered ground state is observed \cite{qcp}. An
obvious question concerns the fate of the intriguing physical
properties such as the absence of coherent magnon excitations
associated with the ``Heisenberg quantum critical point'', if a
magnetic field is applied or ``XY''-like exchange anisotropies are
present. It has been known for a long time that critical behaviour
persists in both cases and there is an entire quantum critical
manifold rather than just a point. In this regime a host of exact
results on thermodynamic quantities as well as dynamical correlation
functions is available (see e.g. [\onlinecite{XXZ}] and references
therein). On the other hand if one combines an exchange anisotropy
with a magnetic field that is at an angle to the anisotropy axis,
altogether different physics emerges \cite{KurmannPA112}. The
transverse field breaks the continuous U(1) symmetry of rotations
around the anisotropy axis and the ground-state develops long-range
N\'eel order. Another interesting feature is that the transversality
of the field prevents the uniform magnetization from reaching
saturation except for infinitely strong fields.

The main motivation for our work are recent Neutron scattering
experiments on the quasi one dimensional antiferromagnet
$\mbox{Cs}_2\mbox{CoCl}_4$ \cite{KenzelmannPRB65} in the presence of a 
magnetic field. It was suggested in Ref.[\onlinecite{KenzelmannPRB65}]
that an appropriate starting point for a description of these
experiments is the anisotropic spin$-1/2$ Heisenberg chain in
transverse magnetic field
\begin{eqnarray}
{\cal H} = J \sum_j S^x_j S^x_{j+1} + \Delta S^y_j S^y_{j+1} +
S^z_j S^z_{j+1} +HS^z_j,
\label{Hamil}
\end{eqnarray}
with $\Delta \approx 0.25$. In view of our interest towards
$\mbox{Cs}_2\mbox{CoCl}_4$, we will often consider $\Delta =
0.25$ in what follows, but our results are more general. The inelastic
Neutron scattering intensity is proportional to
\be
I(\omega,{\bf k})\propto \sum_{\alpha,\beta} (1-\frac{k_\alpha k_\beta}{{\bf k}^2}
)\ S^{\alpha\beta}(\omega,{\bf k})\ ,
\label{intensity}
\ee
where $\alpha, \beta = x,y,z$ and the dynamical structure factor
$S^{\alpha\beta}$ is defined by
\begin{eqnarray}
S^{\alpha \beta} (\omega,{\bf k}) &=& \frac{1}{2\pi N} \sum_{j,l= 1}^N
\int_{-\infty}^{\infty} 
dt\ e^{-i {\bf k}\cdot {\bf R}_l + i \omega t}\nn
&&\qquad\qquad\times  \langle S^{\alpha}_j (t) S^{\beta}_{j-l} (0)
\rangle_c.
\end{eqnarray}
Here the subindex denotes the connected part of the correlator. For a
three dimensional system of uncoupled one dimensional chains we have
${\bf R}_l=l a_0\ {\bf e}$, where ${\bf e}$ is a unit vector pointing
along the chain direction. The dynamical structure factor then depends
only on the component of the momentum transfer along ${\bf e}$, which
we will denote by $k$.

The outline of this paper is as follows: in section \ref{sec:mf} we
summarize essential steps of the mean-field approximation of
Dmitriev {\it et al}. We then discuss the region of applied fields in
which we believe the mean-field approximation to be applicable. In 
section \ref{sec:dsf} we use exact methods to determine all non-zero
components of the dynamical structure factor within the framework of
the mean-field approximation.
In section \ref{sec:SGM} we derive exact results for the structure
factor in the small anisotropy limit by employing field theory
methods. In section \ref{weakfield} we discuss a field theory approach
to the weak field limit. Section \ref{sec:disc} contains a discussion
of our results in the context of the Neutron scattering experiments on
${\rm Cs_2CoCl_4}$.

\section{Mean-Field Approximation}
\label{sec:mf}
\begin{figure}
\psfrag{W}[][]{$\omega/J$}
\psfrag{k}[][]{$k$}
\includegraphics[width=7cm]{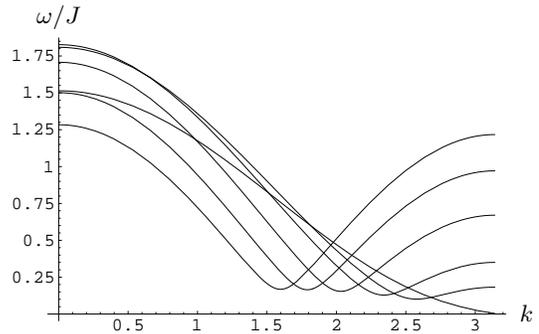}
\caption{Dispersion relation for $\Delta = 0.25$ and external magnetic
fields $H = 0.05, 0.2, 0.4, 0.8, 1.2$ and $1.6$ (top to bottom at $k =
\pi$.}
\label{DispRel}
\end{figure}

Let us first review the main ingredients of the mean-field
approximation (MFA) proposed by Dmitriev {\it et al}. in
Ref. [\onlinecite{dmitriev}]. We note that the MFA
can be applied to the XYZ chain in a magnetic field as
well. One first performs a Jordan-Wigner transformation, which yields
the spinless fermion Hamiltonian 
\begin{eqnarray}
{\cal H}_0 = \sum_j \frac{J_+}{2} (c_j^{\dagger} c_{j+1} + h.c.)
+ \frac{J_-}{2} (c_j^{\dagger} c_{j+1}^{\dagger} + h.c.) 
+ \nonumber \\
+ (n_j - 1/2) (n_{j+1} -1/2) + H (n_j
-1/2).
\end{eqnarray}
Here the hopping and pairing matrix elements are $J_{\pm} = (1 \pm
\Delta)/2$ and we have put $J=1$.  
The pairing term is a consequence of the anisotropy of 
the model:  it vanishes at the isotropic point $\Delta=1$. In the
MFA the four-fermion interaction term is
decoupled in all possible ways. The relevant expectation values
for the on-site magnetization, pairing and kinetic term are denoted by
\begin{eqnarray}
{\cal M} = \langle n_j \rangle -1/2, \hspace{0.5cm}
P = \langle c_{j+1} c_j \rangle, \hspace{0.5cm}
K = \langle c_{j+1}^{\dagger} c_j \rangle .
\end{eqnarray}
The mean-field Hamiltonian is then
\begin{eqnarray}
{\cal H}_{\rm MF} &=&\sum_j\frac{\tilde{J}_+}{2} c_j^{\dagger} c_{j+1} +
\frac{\tilde{J}_-}{2} c_j^{\dagger} c_{j+1}^{\dagger} + h.c.\nn
&&+ \tilde{H}\sum_j (n_j-1/2),
\end{eqnarray}
where we have dropped an unimportant constant, and defined
the (real) parameters
\begin{eqnarray}
\tilde{J}_+ = J_+ - 2K, \hspace{0.5cm}
\tilde{J}_- = J_- + 2P, \hspace{0.5cm}
\tilde{H} = H + 2{\cal M}.
\end{eqnarray}
The theory is now purely quadratic in fermion operators, and can thus
be solved exactly.  An alternative picture of this theory is obtained by
Jordan-Wigner transforming back to spin variables, yielding an
anisotropic $xy$ model in a field,
\begin{eqnarray}
\frac{{\cal H}_{\rm xy}}{\tilde{J}_+} \!=\! \sum_j \left[ (1\!+\!\gamma) S_j^x
S_{j+1}^x + 
(1\!-\!\gamma) S_j^y S_{j+1}^y + h S_j^z \right]
\label{MFH}
\end{eqnarray}
with $\gamma = \tilde{J}_-/\tilde{J}_+$ and $h = \tilde{H}/\tilde{J}_+$.
This model has been extensively studied in the literature and many
useful results on correlation functions are available
\cite{NiemeijerPhysica36,McCoyPRA4,JohnsonPRA4,tracy,TaylorPRB28}.
For our purposes, we first need to complete the mean-field setup above
by providing the necessary self-consistency conditions.  First, we use
the fact that the free fermionic theory can be Fourier transformed and
diagonalized using a Bogoliubov-de Gennes (BdG) transformation, which
reads explicitly
\begin{eqnarray}
c_k = \alpha_+ (k) c'_k + i \alpha_- (k) {c'}^{\dagger}_{-k}, 
\nonumber \\
c^{\dagger}_{-k} = -i \alpha_+ (k) c'_k - \alpha_+ (k) {c'}^{\dagger}_{-k}
\end{eqnarray}
with parameters
\begin{eqnarray}
\alpha_{\pm} (k) = \frac{1}{\sqrt{2}} \left[ 1 \pm \frac{
\tilde{J}_+ \cos k + \tilde{H}}
{\omega(k)} \right]^{1/2}.
\end{eqnarray}
In the thermodynamic limit, the Hamiltonian becomes  
\begin{eqnarray}
\frac{{\cal H}_{\rm MF}}{N} = 
\int_0^{\pi} \frac{dk}{2\pi} {{\bf c}'}^{\dagger} (k) \left( 
\begin{array}{cc}
\omega_+(k) & 0 \\
0 & \omega_-(k) 
\end{array} \right) {\bf c}'(k),
\end{eqnarray}
in which the Nambu spinors have bands $\omega_{\pm} (k) =
\pm \omega (k)$,  
\begin{eqnarray}
\omega (k) = \tilde{J}_+ \sqrt{(\cos k + h)^2 + \gamma^2 \sin^2k}.
\label{disprel}
\end{eqnarray}
This dispersion relation is plotted in Fig.\ref{DispRel} for
$\Delta=0.25$ and various values of the external field. For
$\Delta=0.25$ the gap vanishes for a critical field of $H_c = 1.604$
(corresponding  to $h=1$). The self-consistency conditions at $T=0$
can be cast in the form 
\begin{eqnarray}
{\cal M}_i = \int_0^{\pi} \frac{dk}{2\pi} 
\frac{\partial \omega_- (k)}{\partial \tilde{H}_i},
\end{eqnarray}
where we have defined the shorthand notations 
${\cal M}_0 = {\cal M}, {\cal M}_1 = K, {\cal M}_2 = P, \tilde{H}_0 = \tilde{H},
\tilde{H}_1 = \tilde{J}_+, \tilde{H}_2 = \tilde{J}_-$.  The
numerical solution of these three coupled equations determines
all the necessary mean-field parameters for given values of the
anisotropy and of the external magnetic field.  

\subsection{Mean-Field Phase Diagram}
\label{ssec:MFPD}
The mean-field phase diagram has already been discussed in
Ref.[\onlinecite{dmitriev}].
\begin{figure}[ht]
\begin{center}
\epsfxsize=0.4\textwidth
\epsfbox{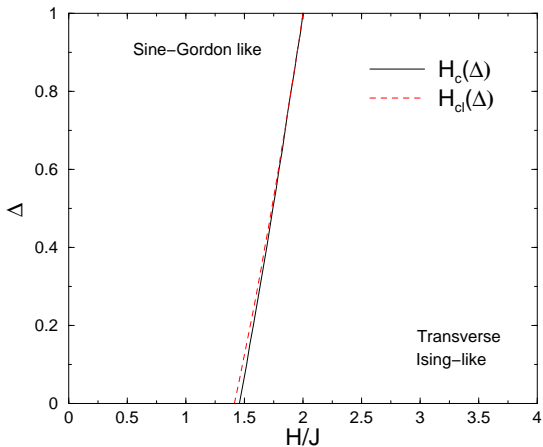}
\end{center}
\caption{\label{Phasediag}
Mean-field phase diagram. The critical line is denoted by
  $H_c(\Delta)$ and the classical line by $H_{cl}(\Delta)$.}
\end{figure}
Here we would like to add some observations concerning the scaling
limits $\gamma\to 0$ and $\gamma\to 1$ respectively. For small
$\gamma$ and $h< 1$ we may bosonize the mean-field Hamiltonian
\r{MFH} and obtain a Quantum Sine-Gordon model, described by the
Hamiltonian density
\bea
{\cal H}=\frac{v}{2}\left[\left(\partial_x\Theta\right)^2
+\left(\partial_x\Phi\right)^2\right]-\gamma\Lambda\cos\sqrt{4\pi}\Theta\ ,
\label{SGM}
\eea
where $\Phi$ is a canonical Bose field and $\Theta$ its dual field,
$\Lambda$ is a constant and $v=\tilde{J}_+\sqrt{1-h^2}$ is
the mean-field spin velocity at $\gamma=0$. For $h>1$ the $\gamma=0$
mean-field Hamiltonian has a ferromagnetic ground state and 
an excitation gap and bosonization cannot be carried out.
The physics of the Sine-Gordon model \r{SGM} is very well
understood. There are two elementary excitations known as soliton and
antisoliton. The spectrum consists of scattering states of an even
number of solitons and antisolitons. By solving the mean-field
equations numerically we find that small $\gamma$ is obtained for
small fields $H$ and goes to zero as $\Delta\to 1$ from below. This
suggests that the physics of the mean-field theory at small fields is
approximately described by the SGM. We note that for $\Delta$ slightly
smaller than $1$ it was shown in Ref.[\onlinecite{fab}] that the full
Hamiltonian \r{Hamil} maps onto a Sine-Gordon model. We elaborate on
this fact in section \ref{sec:SGM}. For $\gamma\approx 1$ it is
convenient to rewrite the Hamiltonian in terms of the real, fermionic
operators
\bea
\eta_n=\sigma^x_n\prod_{j=1}^{n-1}\sigma^z_j\ ,\qquad
\zeta_n=-i\prod_{j=1}^{n}\sigma^z_j \sigma^x_n.
\eea
Then one performs a rotation
\be
\xi_{R,n}=\frac{\zeta_n+\eta_n}{\sqrt{2}},\quad
\xi_{L,n}=\frac{\zeta_n-\eta_n}{\sqrt{2}},
\ee
and finally takes the continuum limit
$\xi_{\alpha,n}\rightarrow\sqrt{a_0}\xi_\alpha(x)$. In this way
one finds that the Hamiltonian \r{MFH} reduces to
\bea
{\cal H}&=&-i\frac{v}{2}\int dx\left[\xi_R\partial_x\xi_R-
\xi_L\partial_x\xi_L\right]-im\int dx\ \xi_R\xi_L,\nn
\label{TFIM}
\eea
where $\xi_{R,L}$ are right and left moving real (Majorana) fermions
respectively and $v=\frac{\gamma\tilde{J}_+a_0}{2}$,
$m=\frac{\tilde{J}_+}{2}(h-1)$. The Majorana fermion theory
\r{TFIM} is the same as the continuum limit of the transverse field
Ising model (TFIM), i.e. the mean-field Hamiltonian for $\gamma=1$,
with a renormalized velocity. The elementary excitations are $Z_2$
kinks. The region $h<1$ corresponds to the ordered phase, whereas
$h>1$ corresponds to the disordered phase of the TFIM. In the ordered
phase the spin operator $\sigma(x)$ has a nonzero expectation value,
which corresponds to the staggered magnetization in the underlying
Heisenberg model. By solving the mean-field equations numerically we
find that $\gamma\approx 1$ is only achieved for small $\Delta\approx
0$ and fields $H$ that are significantly larger than $H_c$.

\subsection{Staggered Magnetization}
Order in the system is represented by the staggered magnetization
$m_{\rm st}=\langle (-1)^j S^x_j \rangle$, which is obtained from
Ref.[\onlinecite{BarouchPRA3}] as  
\begin{eqnarray}
m_{\rm st}
 = \left\{ 
\begin{array}{cc}
\frac{(\tilde{J}_+\tilde{J}_-)^{1/4}\left[ 1 -
\tilde{H}^2/\tilde{J}_+^2 \right]^{1/8}}{
(2(\tilde{J}_+ + \tilde{J}_-))^{1/2}}
& |\tilde{H}| < \tilde{J}_+,
\\
0 & |\tilde{H}| > \tilde{J}_+.
\end{array}
\right.
\label{mfstag}
\end{eqnarray}
\begin{figure}[ht]
\begin{center}
\epsfxsize=0.4\textwidth
\epsfbox{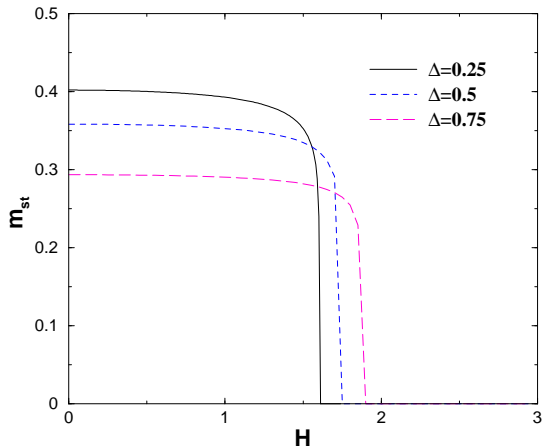}
\end{center}
\caption{Staggered magnetization as a function of external field for
different anisotropies $\Delta = 0.25, 0.5, 0.75$.}
\label{fig:mfstag}
\end{figure}
The mean-field result \r{mfstag} becomes exact on the
classical line $h_{cl}=\sqrt{2(1+\Delta)}$, where 
 \cite{shrock} $m_{\rm st}=\sqrt{1-\frac{h_{cl}^2}{4}}$.
We plot $m_{\rm st}$ as a function of magnetic field for several
values of $\Delta$ in Fig.\ref{fig:mfstag}. 
\subsection{Applicability of the Mean Field Approximation}
\label{ssec:appl}
It is immediately clear from the results for the staggered
magnetization shown in Fig.\ref{fig:mfstag} that the mean-field
theory fails to describe the model \r{Hamil} in the limit of
small applied magnetic fields $H$. In zero field \r{Hamil} reduces to
a critical XZX spin chain, for which the staggered magnetization
vanishes. The XZX chain has a U(1) symmetry corresponding to rotations
around the y-axis, whereas the MFA breaks this
symmetry and is therefore invalid.
Simple scaling arguments suggest that in the limit $H\to 0$,
$m_{\rm st}$ should be proportional to $H^\alpha$, where the exponent
$\alpha$ is a function of the anisotropy $\Delta$ \cite{dmitriev}.
However, a renormalisation group analysis of the small $H$ regime of
\r{Hamil} is nontrivial because in a bosonized description the
transverse field carries conformal spin. We review some important
features of this analysis in section \ref{weakfield}.

In order to get some rough measure concerning the region of magnetic
fields $H$ in which the MFA may work well, we
have computed the one-loop corrections to the mean-field values of
values ${\cal M}$, $K$, $P$. As there is no small parameter in the problem,
this calculation should be understood as a semiclassical
approximation. The corrections to ${\cal M}$, $K$ and $P$ are obtained by
taking partial derivatives of the correction to the ground state
energy. The latter is found to be 
\begin{widetext}
\begin{eqnarray}
\delta E_0 \!= \!\int_{-\pi}^{\pi} \!\!\frac{\rm dp dq dQ}{(2\pi)^3}
\frac{\Bigl[ \frac{1}{2} F(p\!+\!Q,p) F(q\!+\!Q,q) \cos Q  - F(p,q)
F(p\!+\!Q, q\!+\!Q) \cos
(p\!-\!q) \Bigr] F(p\!+\!Q,
p) F(q\!+\!Q, q) \cos Q}{\omega(p) + \omega(q) + \omega(p+Q)+
\omega(q+Q)} 
\nonumber
\end{eqnarray}
\end{widetext}
where
$F(p, q) = \alpha_+ (p) \alpha_- (q) + \alpha_- (p) \alpha_+
(q)$. The corrections to the mean-field parameters ${\cal M}$, $K$, $P$ turn
out to be quite small except for low fields and near the transition:
as the gap becomes smaller, fluctuations become more
important. 

A better way to study the regime of applicability of the mean-field
approximation is to compare some of its predictions to the results of 
numerical Density Matrix Renormalisation Group (DMRG) calculations.
Results for some values of $\Delta$ have been already been reported by
Capraro and Gros in Ref. [\onlinecite{capraro}].

\begin{figure}[ht]
\begin{center}
\epsfxsize=0.45\textwidth
\epsfbox{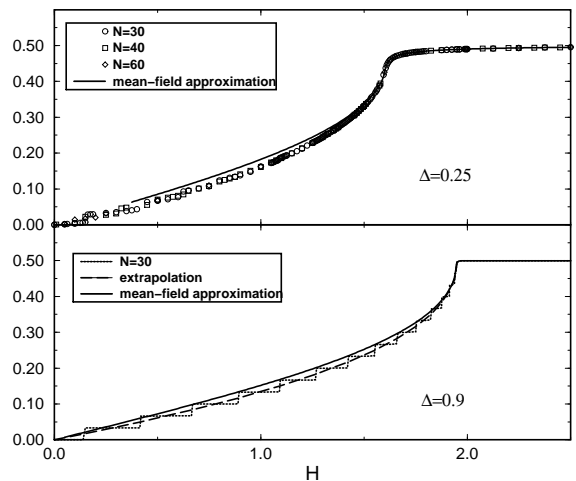}
\end{center}
\caption{\label{fig:magnet}
Magnetisation per site calculated from the MFA
compared to the results of DMRG computations for $\Delta=0.25$.}
\end{figure}

In Fig.\ref{fig:magnet} we compare DMRG results for the magnetisation
per site with the MFA for $\Delta=0.25$. We work with periodic
boundary conditions and consider system sizes of up to 60 sites.
The agreement between DMRG and the MFA is quite satisfactory,
particularly for sufficiently large fields $H\agt 1.2J$. 

\begin{figure}[ht]
\begin{center}
\epsfxsize=0.45\textwidth
\epsfbox{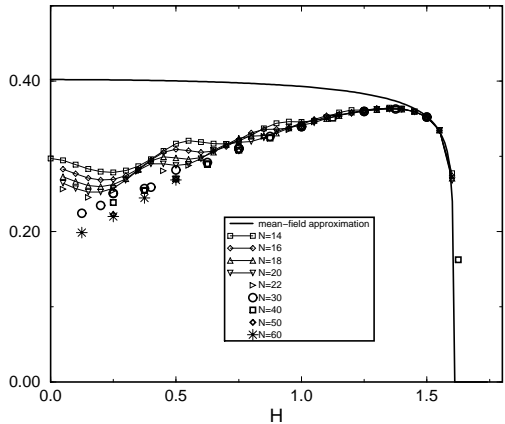}
\end{center}
\caption{\label{fig:stagm}
Staggered magnetisation calculated from the MFA
compared to the results of DMRG computations for $\Delta=0.25$.}
\end{figure}

\begin{figure}[ht]
\begin{center}
\epsfxsize=0.45\textwidth
\epsfbox{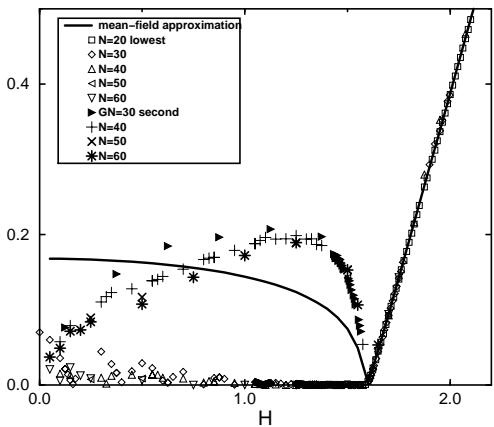}
\end{center}
\caption{\label{fig:gap}
Excitation gap calculated from the MFA
compared to the results of DMRG computations for $\Delta=0.25$.}
\end{figure}

In Fig.\ref{fig:stagm} we compare DMRG results for the staggered
magnetisation per site with the MFA for
$\Delta=0.25$. For $H<H_c$ the staggered magnetization is computed as
\be
\langle 0| S^x_n(-1)^n|1\rangle\ ,
\label{overlap}
\ee
where $|0\rangle$ and $|1\rangle$ are the two degenerate ground states
corresponding to momenta $0$ and $\pi$ respectively. It is necessary
to consider the overlap \r{overlap}, because in the finite volume and
periodic boundary conditions, translational invariance implies that
\be
\langle 0| S^x_n(-1)^n|0\rangle = \langle 1| S^x_n(-1)^n|1\rangle = 0\ .
\ee
In the infinite volume translational invariance is spontaneously
broken and the ground state becomes a linear combination of
$|0\rangle$ and $|1\rangle$. We note that the finite-size effects
for weak and strong fields are still rather pronounced. Such behaviour
for the staggered magnetization in finite systems has been previously
observed in e.g. Ref. [\onlinecite{Fabr91}]. Nevertheless it is clear
that the staggered magnetization vanishes above a critical field and
approaches zero in the weak field limit. The agreement between the
DMRG results and the MFA for weak fields is as
expected quite poor. For sufficiently large field $H\agt 1.2J$ the
MFA gives good results.
Last but not least we have used DMRG to determine the gaps of the two
lowest lying excited states. The results are shown in
Fig. \ref{fig:gap}. Both gaps vahish at the critical field $H_c$.
For $H<H_c$ the first excited state is really a degenerate ground
state with the opposite sign of the staggered magnetization. This
degenereacy is removed in the thermodynamic limit by spontaneous
symmetry breaking. Hence the true excitation gap is given by the
second excited state. 
We see that the MFA is in very good agreement
with the DMRG results for fields that are larger than approximately
$1.5J$. This is only slightly below the critical field $H_c$.

The above considerations suggest that the MFA
works very well for both ground state and excited states properties as
long as the applied field is sufficiently strong $H\agt 1.5J$. For
intermediate field strengths $0.5J\alt H\alt 1.5J$ we expect the
MFA to give at least qualitatively correct
results. In the small-field regime the MFA gives very poor results and
fails complately in the limit $H\to 0$. Having in mind these
limitations we will now determine the dynamical structure factor in
the MFA.

\section{Dynamical Structure Factor in the Mean-Field Approximation}
\label{sec:dsf}

The transverse correlators are determined in the framework of the
mean-field theory as follows. Let us denote the true ground state by
$|0\rangle$ and the mean-field ground state by $|0\rangle_{\rm MF}$.
Then we employ the following approximation
\bea
&&\langle 0|S^\alpha_n(t)\ S^\beta_m|0\rangle=
\langle 0|e^{iH_0t}S^\alpha_ne^{-iH_0t}\ S^\beta_m|0\rangle\nn
&&\quad\qquad\approx
{_{\rm MF}\langle} 0|e^{iH_{\rm xy}t}S^\alpha_ne^{-iH_{\rm xy}t}\
S^\beta_m|0\rangle_{\rm MF}.
\label{mfapproxSS}
\eea
In other words we replace the ground state expectation value by the
expectation value with respect to the mean-field ground state and 
substitute the xy-Hamiltonian \r{MFH} for the full Hamiltonian in the
time evolution operator. It is clear that
\r{mfapproxSS} will be a good approximation as long as the mean-field
theory gives an accurate description of the ground state and low-lying
excited states. The transverse spin correlators in the theory \r{MFH}
have been determined as a spectral sum over intermediate multiparticle
states in Refs [\onlinecite{McCoyPRA4,JohnsonPRA4,tracy}].  

\subsection{Transverse correlations in the ``Low-field Phase'': $h<1$}

For $h < 1$, the leading contribution comes from two-particle
intermediate states. As a result all components of the  structure
factor are completely incoherent. 
The other contributions are due to states with $4,6,...$ particles,
but they are expected to contribute less spectral weight so that we
ignore them here. The two-particle contributions to the structure
factor are given by
\bea
S^{\alpha\beta}(\omega,k) = \int_{-\pi}^\pi \frac{dk_1\ dk_2}{2\pi}
\delta(\omega-\omega(k_1)-\omega(k_2))\ \times \nonumber \\
\times \delta(k-k_1-k_2 + \pi)\
g_{\alpha\beta}(k_1,k_2)\ .
\eea
Here
\bea
g_{xx}(k_1,k_2)&=&\frac{C (1+\gamma)^{-1}}{8
[1-\cos(k_1+k_2)]}\frac{\left[\omega(k_1)-\omega(k_2)\right]^2}{\omega(k_1)\omega(k_2)}\
,\nn 
g_{yy}(k_1,k_2)&=&
\frac{C (1+\gamma)}{8}
\frac{1-\cos(k_1-k_2)}{\omega(k_1)\omega(k_2)}\ \tilde{J}_+^2,\nn 
g_{xy}(k_1,k_2)&=&i\frac{C}{8}
\frac{\sin\left(\frac{k_1-k_2}{2}\right)}{\sin\left(\frac{k_1+k_2}{2}\right)}
\left[\frac{\tilde{J}_+}{\omega(k_1)}-\frac{\tilde{J}_+}{\omega(k_2)}\right],
\eea
where $C = \left[\gamma^2 (1-h^2) \right]^{\frac{1}{4}}$ and where we
have corrected a typo in Ref.\onlinecite{tracy}. The result for $g_{xy}$ is a
conjecture based on calculations we have performed in the Ising and
Sine-Gordon scaling limits and agrees with the known result on the
classical line \cite{shrock}. We note that the results for
$S^{\alpha\alpha}(\omega,k)$ reduce to the exact results \cite{shrock}
for the model \r{MFH} on the classical line as well, which implies
that all multiparticle matrix elements vanish. We see that the
two-particle contribution to the off-diagonal element $S^{xy}(\omega,k)$ is purely
imaginary. This means that it cannot be seen in Neutron scattering
experiments as it enters the expression \r{intensity} for the intensity in
the form
\be
(1-\frac{k_xk_y}{{\bf k}^2})\left[
  S^{xy}(\omega,k)+S^{yx}(\omega,k)\right], 
\ee
where we recall that $k={\bf k}\cdot {\bf e}$ is the component of the
momentum transfer along the chain direction. Using the relation
\be
S^{yx}(\omega,k)=\left(S^{xy}(\omega,k)\right)^*\ ,
\ee
we see that $S^{xy}(\omega,q)+S^{yx}(\omega,q)$ is zero whenever
$S^{xy}$ is purely imaginary.

\begin{figure}
\includegraphics[width=8cm]{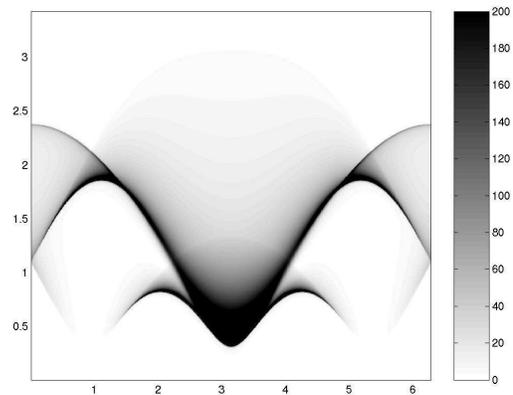}
\caption{Structure factor $S^{xx}$ as a function of $k$ and $\omega$
for anisotropy $\Delta = 0.25$ and external magnetic field $H = 0.8J$.}
\label{SFXX.H.0.8}
\end{figure}

\begin{figure}
\includegraphics[width=8cm]{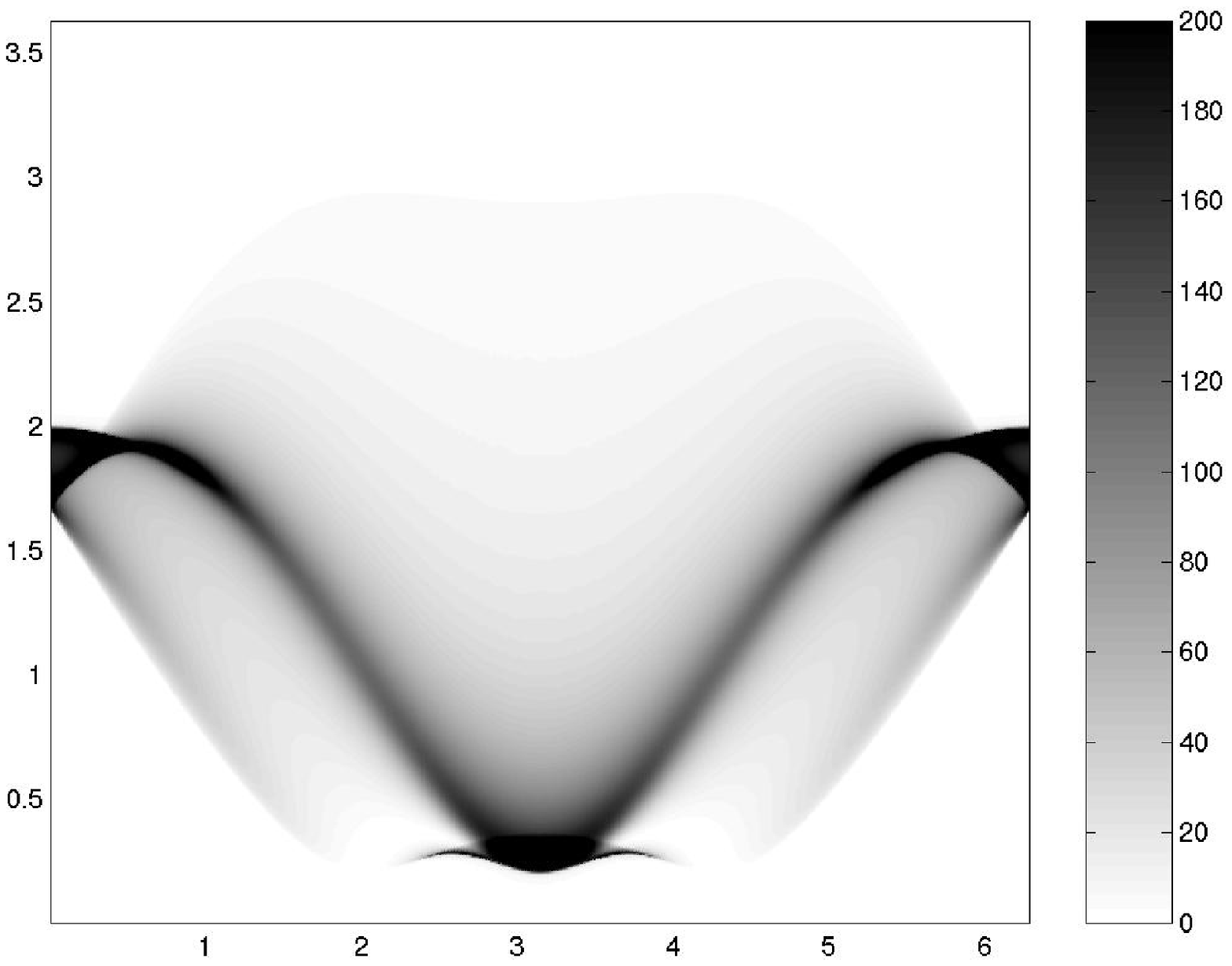}
\caption{Structure factor $S^{xx}$ as a function of $k$ and $\omega$
for anisotropy $\Delta = 0.25$ and external magnetic field $H = 1.4J$.}
\label{SFXX.H.1.4}
\end{figure}

\begin{figure}
\includegraphics[width=8cm]{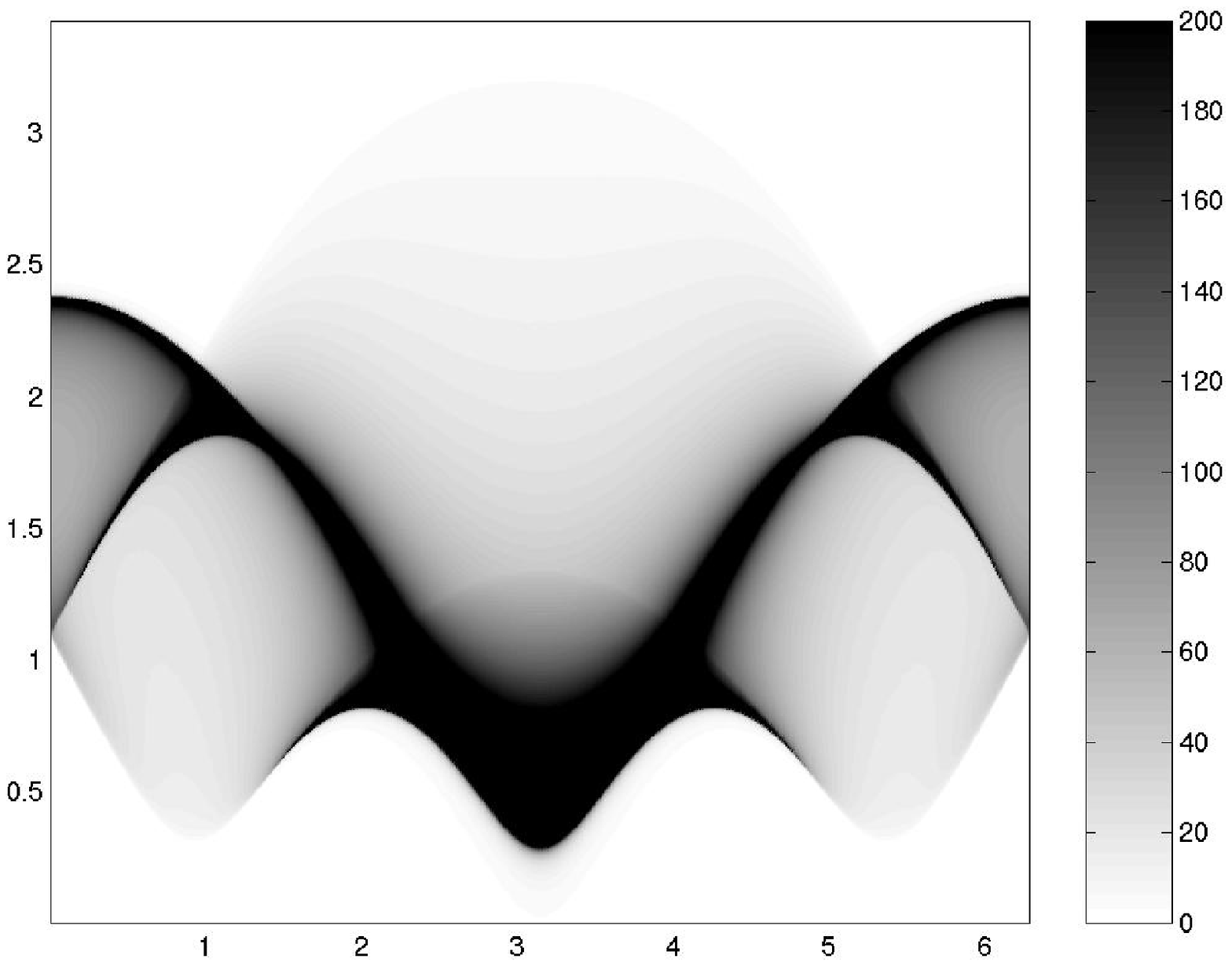}
\caption{Structure factor $S^{yy}$ as a function of $k$ and $\omega$
for anisotropy $\Delta = 0.25$ and external magnetic field $H = 0.8J$.}
\label{SFYY.H.0.8}
\end{figure}

\begin{figure}
\includegraphics[width=8cm]{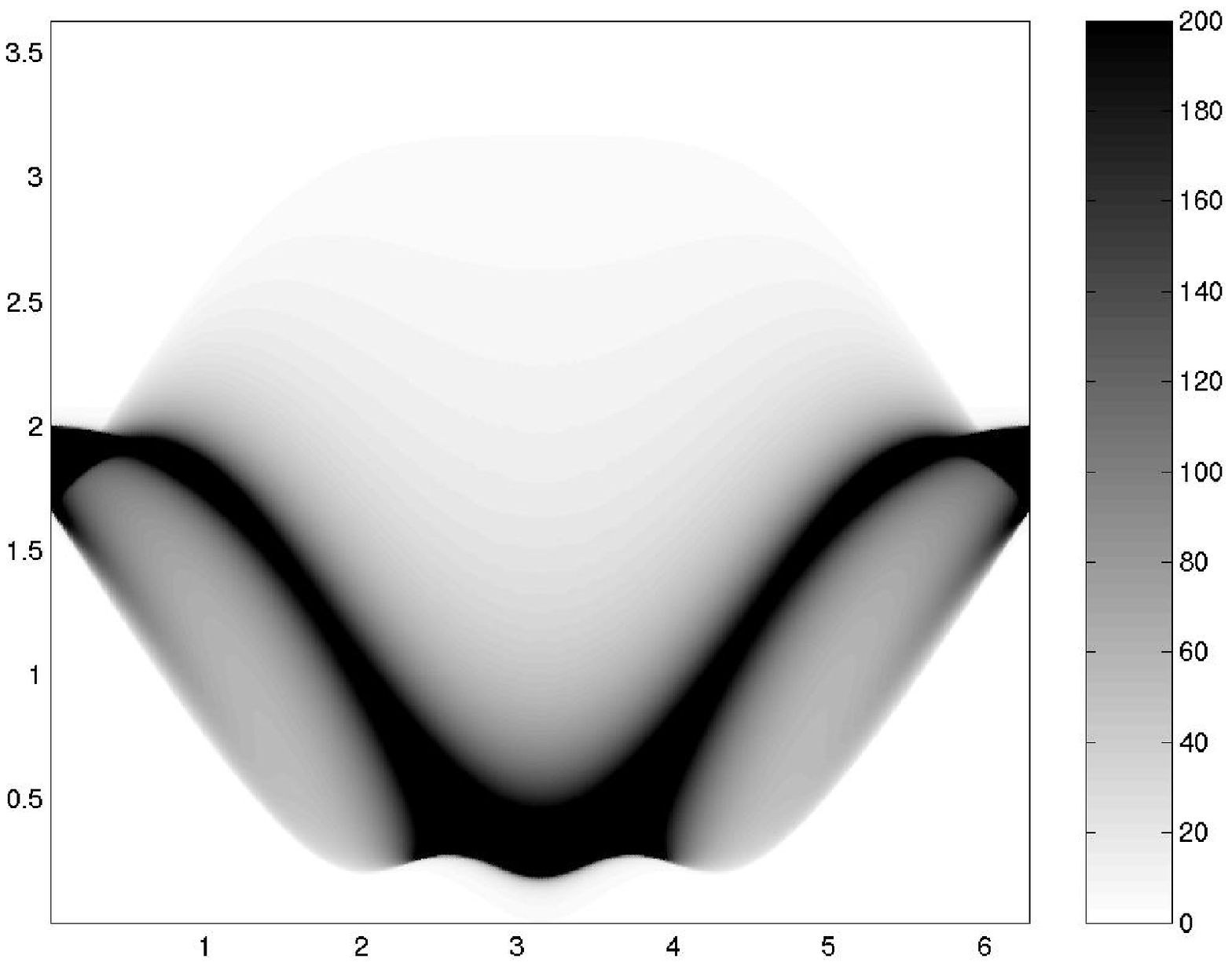}
\caption{Structure factor $S^{yy}$ as a function of $k$ and $\omega$
for anisotropy $\Delta = 0.25$ and external magnetic field $H = 1.4J$.}
\label{SFYY.H.1.4}
\end{figure}

The intensity of the structure factor $S^{\alpha \alpha}$ is plotted as a
function of $k$ and $\omega$ in Figs \ref{SFXX.H.0.8} to
\ref{SFYY.H.1.4} for anisotropy $\Delta = 0.25$ and for external magnetic
fields $H=0.8J$ and $H=1.4J$, corresponding to the effective
mean-field values $h = 0.44$ and $h = 0.82$ respectively. It follows
from our discussion in section \ref{ssec:appl} that we do not expect the
results for $h=0.44$ to provide a quantitative description of the
structure factor of the transverse field XXZ chain. However the
mean-field approximation may still capture the redistribution of
spectral weight in the various components of the structure factor as
the transverse field is increased on a qualitative level.
One clearly sees in {\sl e.g. }  Fig.\ref{SFXX.H.0.8} that the
magnetic field splits the lower boundary into two branches.
Experimentally, one would thus expect to see a double peak in the
intensity when scanning in frequency for fixed momentum.  In Fig.
\ref{SFXX.H.1.4}, the evolution of the structure factor with increasing
magnetic field can be seen:  the branches recollapse around $k=\pi$,
with diminishing gap (the gap vanishes at $H_c = 1.604J$,
corresponding to $h=1$).  As the field gets closer to the critical
field, the intensity of the structure factor collapses from an
incoherent continuum onto an emergent single coherent mode.  

\subsection{Transverse correlations in the ``High-field Phase'': $h>1$}

For fields $h>1$ the dominant contribution to the dynamical structure
factor is due to a single-particle coherent mode with a dispersion
relation given by (\ref{disprel}). We have
\cite{McCoyPRA4,JohnsonPRA4,tracy}
\bea
S^{\alpha\beta}(\omega,k)&=&f_{\alpha\beta}(k-\pi)\ \delta(\omega-\omega(k))\ .
\eea
where
\bea
f_{xx}(k)= \frac{\rho_\infty}{[A(k)]}, \hspace{1cm}
f_{yy}(k)=\rho_\infty\ A(k),\nn 
f_{xy}(k)=-i\rho_\infty, \hspace{1cm}
f_{xz}(k)=f_{yz}(k)=0\ .
\eea
Here we have introduced the parameters
\bea
\lambda_{1,2}&=&\frac{h\pm\sqrt{h^2+\gamma^2-1}}{1-\gamma}\ ,\nn
\rho_\infty&=&\frac{1}{4}\left[(1-\lambda_2^2)(1-\lambda_1^{-2})(1-\lambda_2\lambda_1^{-1})
\right]^\frac{1}{4}\ ,\nn 
A(k)&=&\left[\frac{\lambda_2^2-2\lambda_2\cos(k)+1}{\lambda_1^{-2}-2\lambda_1^{-1}\cos(k)+1}
\right]^{\frac{1}{2}}\ .
\eea
We note that $f_{xy}$ is purely imaginary, which again means that the
mixed correlations cannot be observed in Neutron scattering experiments.

\begin{figure}
\psfrag{SFXX}[][]{$f^{xx}$}
\psfrag{k}[][]{$k$}
\includegraphics[width=7cm]{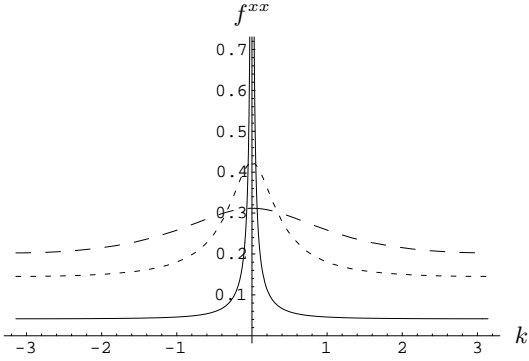}
\caption{$f^{xx}$ structure factor amplitude as a function of $k$ for
external magnetic fields $H = 1.604, 1.7$ and $2.4$ (above the
critical field). This is the amplitude of the structure factor along
the dispersion relation $\delta$-function.}
\label{SFXX3}
\end{figure}

Near the transition, the amplitude $f_{xx}$  diverges at $k=\pi$. For
fields higher than the critical field, the divergence is smoothed out,
as can be seen in figure (\ref{SFXX3}).  The gap also reopens, as can
be seen from equation (\ref{disprel}). 

\subsection{Longitudinal correlations}

The longitudinal correlaion function $S^{zz}$ can be computed directly
in terms of a density-density correlator of the $c$ fermions. After
the BdG transformation to $c'$ fermions, a straightforward vacuum
expectation value yields \cite{TaylorPRB28} 
\begin{eqnarray}
S^{zz}(\omega,k) &=& \!\int_0^{\pi} \frac{dq}{4\pi} [1\!-\!f(q,k)]\
\delta (\omega -\omega(q,k))\ ,\nn
f (q+\frac{k}{2},k) &=& \frac{a(q) a(q+k) - b(q) b(q+k)}
{\omega (q) \omega(q+k)}\ ,\nn
\omega(q,k)&=&\omega (q-k/2) - \omega (q\!+\!k/2)).
\label{szzmf}
\end{eqnarray}
where $a(k)=\tilde{J}_+ \cos(k)+\tilde{H}$ and $b(k) = \tilde{J}_-
\sin (k)$. Equation \r{szzmf} is valid in both the low-field and in
the high-field phase. As a result, the longitudinal correlations are
incoherent for any value of the applied magnetic field.

\begin{figure}
\includegraphics[width=8cm]{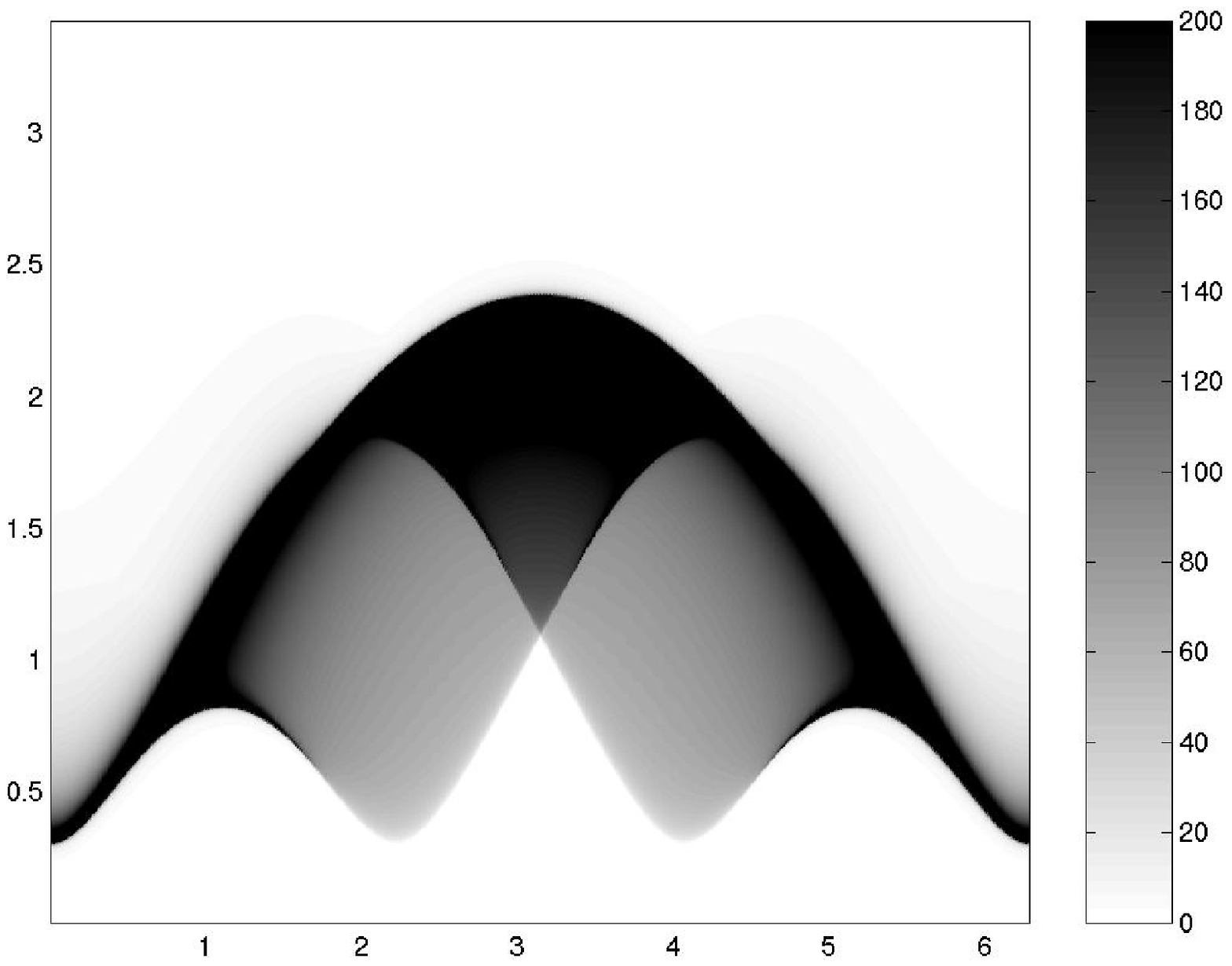}
\caption{Structure factor $S^{zz}$ as a function of $k$ and $\omega$
for anisotropy $\Delta = 0.25$ and external magnetic field $H = 0.8J$.}
\label{SFZZ.H.0.8}
\end{figure}

\begin{figure}
\includegraphics[width=8cm]{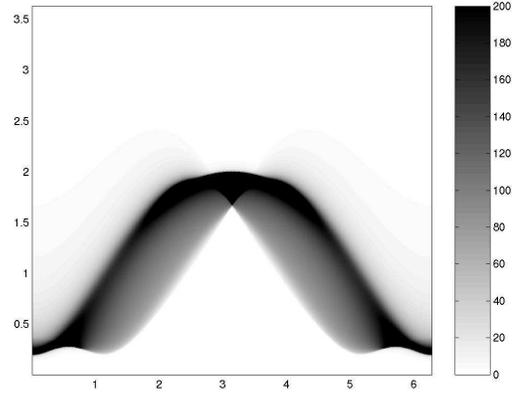}
\caption{Structure factor $S^{zz}$ as a function of $k$ and $\omega$
for anisotropy $\Delta = 0.25$ and external magnetic field $H = 1.4J$.}
\label{SFZZ.H.1.4}
\end{figure}
The evolution of $S^{zz}$ with increasing applied field $H$ is shown
in Figs \ref{SFZZ.H.0.8}-\ref{SFZZ.H.1.4}. Firstly, the longitudinal
correlations are generally incommensurate. Second, in low fields the
intensity is concentrated at the upper boundary of the continuum, as
is expected as $\Delta$ is small, whereas for larger fields a shift
towards lower energies is visible.

\section{Field Theory Approach to the small-anisotropy limit}
\label{sec:SGM}

The easiest case to deal with by field theory methods is the one where
the magnetic field $H$ is much stronger than the anisotropy
$1-\Delta$. The field theory limit in this case has been studied in
Ref.[\onlinecite{fab}] and subsequently in Ref.[\onlinecite{esr}]. Let
us consider the Hamiltonian 
\bea
{\cal H}_{\rm ZXX,H}&=&J\sum_{j,\alpha}
S^\alpha_jS^\alpha_{j+1}+H\sum_j S^z_j\nn
&&+(\Delta-1) S^y_jS^y_{j+1}\equiv{\cal H}_0+{\cal H}_1 .
\label{hxxz}
\eea
As we assume the field $H$ to be much larger than the anisotropy
$1-\Delta$, we bosonize at the point $\Delta=1$ in the presence of a
strong field and then switch on the exchange anisotropy as a
perturbation.

At low energies ${\cal H}_0$ is described by a free massless boson
compactified on a ring of radius $R$ i.e. $\Phi$ and
$\Phi+2\pi R$ are identified. The dual field $\Theta$ fulfils
$\Theta=\Theta+1/R$. The bosonization rules are:
\be
\vec{S}_n \longrightarrow \vec{ J}(x) + (-1)^n\vec{ n} (x) +\ {\rm
  higher\ harmonics},
\ee
where $a_0$ is the lattice spacing, $x=na_0$ and
\bea
J^x&=& b\cos(\beta\Theta(x))\
\sin\left(\frac{2\pi}{\beta}\Phi(x)-2\delta x\right),\nn 
J^y&=& -b\sin(\beta\Theta(x))\
\sin\left(\frac{2\pi}{\beta}\Phi(x)-2\delta x\right),\nn 
J^z&=&\frac{a_0}{\beta}\partial_x\Phi(x)\ ,\nn 
n^x (x) &=& c \cos(\beta\Theta (x)),\
n^y (x) = c \sin(\beta\Theta (x)),\nn
n^z (x) &=&a \sin(\frac{2\pi}{\beta}\Phi(x)-2\delta x).
\label{boso}
\eea
Here $\beta=2\pi R$ and the coefficients $a,c$ are
known exactly in the absence of a magnetic field \cite{lukyanov} and
numerically for several values of the applied field $H$ \cite{FH,HF2}. The
(magnetic-field dependent) constant $\beta$, the incommensurability
$\delta$ and spin velocity $v_s(H)$ are determined by the microscopic
parameters $\Delta$, $J$ and $H$ of the lattice model by using the
exact Bethe Ansatz solution. This entails solving some linear integral
equations numerically (see e.g. Refs.[\onlinecite{vladb,fab,ET,oa2}]). 
The bosonization formulas \r{boso} hold as long as $H<2J$. For $H>2J$
the ground state of ${\cal H}_0$ is the saturated ferromagnetic state
and the excitation spectrum is gapped. For later convenience we define  
\be
\xi=\frac{4\beta^2}{8\pi-4\beta^2}\ .
\ee
The perturbing Hamiltonian ${\cal H}_1$ can now be bosonized using
\r{boso} and fusion of the staggered magnetizations $n^y$ gives a
contribution proportional to 
\bea
\cos(2\beta\Theta)\ .
\eea
In addition there is a small marginal contribution that shifts the
compactification radius. For simplicity we neglect it here. 

Thus, at low energies compared to the scale set by the applied field
$H<2J$, the effective Hamiltonian is given by a Sine-Gordon model
\bea
{\cal
H}=\frac{v_s}{2}[(\partial_x\Phi)^2+(\partial_x\Theta)^2]
-\mu(\Delta)\ \cos(2\beta\Theta)\ .
\label{SGM2}
\eea
The cosine term in the Sine-Gordon model is relevant and generates a
spectral gap. As $2\beta>\sqrt{4\pi}$ (see e.g. Fig.1 of
Ref.[\onlinecite{ET}]), the spectrum of the Sine-Gordon model consists
of soliton and antisoliton only. We can immediately read off the
scaling of the gap as a function of $1-\Delta$ 
\bea
M\propto (1-\Delta)^\frac{\pi}{2\pi-\beta^2}\ .
\eea
The magnetic field dependence enters both via the prefactor and via
the $H$-dependence of $\beta$. 
In order to calculate the prefactor as well as quantities like the
magnetization we need to know the normalisation of the operator
\bea
{\cal O}_j=S^y_jS^y_{j+1}\longrightarrow -{\cal C}\ \cos(2\beta\Theta)
\label{normalisation}
\eea
in the Heisenberg chain in a field, i.e. the Hamiltonian \r{hxxz} with $\Delta=1$.
At present the normalisation ${\cal C}$ is not known. In
Ref.[\onlinecite{HF2}] the issue of how to determine ${\cal C}$ from the
large-distance asymptotics of appropriately chosen correlation
functions in the Heisenberg chain in a uniform field has been investigated. 
In what follows we will consider ${\cal C}$ as a yet unknown function of the
magnetic field $H$. The gap is given by 
\bea
\frac{M}{J}&=&\frac{2\tilde{v}}{\sqrt{\pi}}
\frac{\Gamma\left(\frac{\eta}{2-2\eta}\right)}
{\Gamma\left(\frac{1}{2-2\eta}\right)}
\left(
\frac{(1-\Delta){\cal C}\pi}{2\tilde{v}}
\frac{\Gamma\left(1-\eta\right)}{\Gamma\left(\eta\right)}
\right)^\frac{1}{2-2\eta}\ ,\nn
\eea
where 
\be
\eta=\beta^2/2\pi\ ,
\ee
and $\tilde{v}$ is the dimensionless spin velocity
\be
\tilde{v}=\frac{v_s}{Ja_0}.
\ee
The staggered magnetization in x-direction is nonzero in the presence
of a transverse field. We have 
\bea
\langle(-1)^n S^x_n\rangle=c\ \langle\cos\beta\Theta\rangle\ ,
\label{SGMSMag}
\eea
where \cite{LZ}
\begin{widetext}
\bea
\langle\cos\beta\Theta\rangle&=&\left[\frac{M\sqrt{\pi}
\Gamma(\frac{1}{2-2\eta})}{2\tilde{v}J\Gamma(\frac{\eta}{2-2\eta})}
\right]^{\frac{\eta}{2}}\exp\left[\int_0^\infty\frac{dt}{t}\left(\frac{\sinh(\eta
    t)}{2\sinh(t)\cosh([1-\eta]t)}-\frac{\eta}{2} e^{-2t}\right)\right].
\eea
\end{widetext}
The magnetization per site can be calculated from the ground state energy
of the Hamiltonian by taking derivatives with respect to the magnetic
field $H$. The total ground state energy per site is given by
\be
e_{\rm tot}(H)=e_{\rm XXX}(H)+e_{\rm SG}(H,\Delta)\ ,
\ee
where the ground state emergy per site of the isotropic Heisenberg
lattice model $e_{\rm XXX}(H)$ is known exactly from the Bethe Ansatz
(see e.g. Ref.\onlinecite{vladb}) and where $e_{\rm SG}$ is the ground
state energy of the Sine-Gordon model \cite{destri} 
\bea
e_{\rm SG}(H,\Delta)&=&-\frac{M^2}{4\tilde{v}J}\ \tan\left(\frac{\pi\xi}{2}\right).
\eea
The magnetization is given by
\bea
m=\frac{\partial e_{\rm tot}(H)}{\partial H}=m_{\rm XXX}
+\frac{\partial e_{\rm SG}(H,\Delta)}{\partial H}.
\label{SGMMag}
\eea
The only unknown in the expression for the staggered magnetization
\r{SGMSMag} and magnetization \r{SGMMag} is the normalization ${\cal
  C}$ in \r{normalisation}. Once this is known with sufficient
accurary for taking derivatives with respect to $H$ both $m$ and
$\langle(-1)^n S^x_n\rangle$ can be evaluated.

\subsection{Dynamical Structure Factor}
\begin{figure}[ht]
\begin{center}
\epsfxsize=0.4\textwidth
\epsfbox{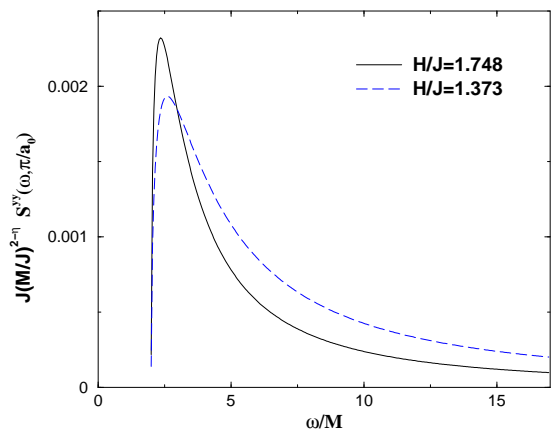}
\end{center}
\caption{\label{fig:syyFT}
$J(M/J)^{2-\eta}S^{yy}(\omega,\frac{\pi}{a_0})$ as a function of
$\omega/M$ for $H=1.373J$ and $H=1.748J$, corresponding to a
  magnetisation per site of $0.2$ and $0.3$ respectively. The scales
  of both $x$ and $y$ axes depends on $H$ through the gap $M$.
}
\end{figure}

Using the integrability of the SGM it is possible to evaluate the
low-energy asymptotics of dynamical correlation functions.
Let us start with the transverse correlations and concentrate on
momenta close to $\pi/a_0$. The staggered magnetizations in $x$ and
$y$ direction are given by \r{boso} and calculating their correlation
functions reduces to the calculation of particular correlators in the
SGM. This is by now a standard calculation (see
e.g. Ref.[\onlinecite{ET}]): one determines the first few terms in a
Lehmann representation by using the exact matrix elements
\cite{smirnov} of the operator under consideration between the ground
state and (multi) soliton/antisoliton excited states. The leading
contribution is due to intermediate states with one soliton and one
antisoliton. Using the notations of Ref.[\onlinecite{lukyanov1}] we
obtain 
\bea
&&S^{yy}(\omega,\frac{\pi}{a_0}+k)=\frac{2\tilde{v}Jc^2}
{\pi s\sqrt{s^2-4M^2}}\left|
\frac{{\cal G}_\beta\ G(2\theta_0)/{\cal C}_1}{\xi\
  \cosh\left(\frac{2\theta_0+i\pi}{2\xi}\right)} \right|^2\nn
&&\hskip 3cm\times\ \Theta_{H}\left(\frac{s}{M}-2\right)\nn
&&\qquad \quad+{\rm contrib.\ from\ 4,6,\ldots\ particles} ,
\eea
where $\Theta_H(x)$ is the Heaviside function, $s$ and $\theta_0$ are
defined as 
\be
s=\sqrt{\omega^2-v_s^2k^2}\ ,\quad
\theta_0={\rm arccosh}\left(\frac{s}{2M}\right),
\ee
and ${\cal G}_\beta$ and $G(\theta)$ are given by
\begin{widetext}
\bea
G(\theta)&=&iC_1\sinh(\theta/2)
\exp\left[\int_0^\infty\frac{dt}{t}\left(
\frac{\sinh^2(t[1-i\theta/\pi])\sinh(t[\xi-1])}{\sinh(2
  t)\sinh(\xi t)\cosh(t)} \right)\right],\nn
{\cal G}_\beta&=&\left[\frac{M\sqrt{\pi}}{2\tilde{v}J}
\frac{\Gamma\left(\frac{1}{2-2\eta}\right)}{\Gamma\left(\frac{\eta}
{2-2\eta}\right)}\right]^{\frac{\eta}{2}}
\exp\left[\int_0^\infty\frac{dt}{t}\left(
\frac{\sinh\eta t}{2\sinh(t)\cosh([1-\eta]t)}
-\frac{\eta}{2} e^{-2t}\right)\right],
\eea
\end{widetext}
where
\bea
{\cal C}_1&=&\exp\left[-\int_0^\infty \frac{dt}{t}
\frac{\sinh^2(t/2)\ \sinh(t[\xi-1])}{\sinh 2t\ \sinh\xi t\ \cosh t}\right].
\label{c1}
\eea
The analogous result for $S^{xx}(\omega,\frac{\pi}{a_0}+k)$ is 
\bea
&&S^{xx}(\omega,\frac{\pi}{a_0}+k)=
\frac{2\tilde{v}Jc^2}{\pi s\sqrt{s^2-4M^2}}\left|
\frac{{\cal G}_\beta\ G(2\theta_0)/{\cal C}_1}{\xi\
  \sinh\left(\frac{2\theta_0+i\pi}{2\xi}\right)} \right|^2\nn
&&\hskip 3cm\times\ \Theta_{H}\left(\frac{s}{M}-2\right)\nn
&&\qquad\quad +\ {\rm contrib.\ from\ 4,6,\ldots\ particles}.
\eea
Finally we determine the longitudinal structure factor
$S^{zz}(\omega,k)$ by making use of
recent results by Lukyanov and Zamolodchikov \cite{lukyanov2}. It is
clear from the bosonization formulas \r{boso} that the correlations of
$n^z$ are incommensurate. In other words the longitudinal structure
factor has low-energy modes at the incommensurate momenta
$\frac{\pi}{a_0}\pm 2\delta$. Moreover one can easily establish that
the operators 
\be
\exp\left(\pm i\frac{2\pi}{\beta}\Phi\right)\ ,
\label{VO}
\ee
have topological charges of $\mp 2$ respectively. Here the operator of
topological charge is defined as
\be
Q=\frac{\beta}{\pi}\int_{-\infty}^\infty dx\ {\partial_x}\Theta(x).
\ee
The leading contribution to the longitudinal spin-spin correlation
functions is due to intermediate states with two solitons or two
antisolitons. 
The formfactors of vertex operators of the kind \r{VO} with definite
topological charge have been given in Ref.[\onlinecite{lukyanov2}]. A
short calculation leads to the following result
\bea
&&S^{zz}(\omega,\frac{\pi}{a_0}\pm 2\delta+k)=
\frac{\tilde{v}Ja^2Z_2(0)}{4\pi s\sqrt{s^2-4M^2}}\left|
G(2\theta_0)\right|^2\nn
&&\hskip 3cm\times\ \Theta_{H}\left(\frac{s}{M}-2\right)\nn
&& +\ {\rm contrib.\ from\ 4,6,\ldots\ particles}.
\eea
The normalization $Z_2(0)$ is
\begin{widetext}
\bea
Z_2(0)
&=&\frac{8}{\xi C_1^2}
\left[\frac{\sqrt{\pi}\ M\Gamma\left(\frac{3}{2}+\frac{\xi}{2}\right)}
{J\tilde{v}\Gamma\left(\frac{\xi}{2}\right)}\right]^\frac{2\pi}{\beta^2}
\exp\left[
\int_0^\infty\frac{dt}{t}\left(\frac{\cosh(t)-\exp(-[1+\xi]t)}
{\sinh(\xi t)\cosh(t)}-\frac{2\pi\ e^{-2t}}{\beta^2}\right)\right],
\eea
\end{widetext}
where $C_1$ is given in \r{c1}.
The structure factor depends on the transverse field both through the
gap $M$ and through the parameter $2\beta$.
The variation of $2\beta$ does not alter the various components of
the dynamical structure factor on a qualitative level: the leading
contributions are always due to two particles, the structure factor is
entirely incoherent and always vanishes at the threshold (as $2\beta$
is always strictly larger than $\sqrt{4\pi}$), which occurs at
$\omega=2M$. However, on a quantitative level the value of $\beta$
is quite important: decreasing $\beta$ leads to a narrowing of the
lineshape in the transverse correlations. What precisely me mean by
this is illustrated in Fig.\ref{fig:syyFT}, where we plot
$J(M/J)^{2-\eta}S^{yy}(\omega,\frac{\pi}{a_0})$ as a function of
$\omega/M$ for several values of the magnetic field $H$.
We see that the lineshape, when measured in units of the
field-dependent gap $M$, sharpens with increasing magnetic field
$H$. However, as $M$ itself depends on $H$, this does not necessarily
imply that the lineshape measured in physical units like ${\rm meV}$
sharpens with increasing $H$. 

It is somewhat difficult to compare our field theory results to the ones
obtained in the mean-field approximation as we have no reliable way of
determining the gap $M$. 

\section{Field theory approach to the weak field limit}
\label{weakfield}
The situation where the field is weak compared to the
anisotropy, i.e. $H\ll (1-\Delta)$, has been analyzed by
renormalisation group methods by Nersesyan {\sl et al} in
Ref. [\onlinecite{shura}]. It is convenient to rotate the quantization
axis such that the anisotropy is in z-direction and the field is
along the x-direction. The effect of the magnetic field is to
generate an excitation gap and a nonzero expectation value for the
staggered magnetization in $y$-direction. This can be seen as
follows. The chain Hamiltonian is 
\bea
{\cal H}_{\rm XXZ,H}&=&J\sum_j S^x_jS^x_{j+1}+S^y_jS^y_{j+1}+\Delta
S^z_jS^z_{j+1}\nn 
&&+H\sum_j S^x_j = {\cal H}_0+{\cal H}_1\ ,
\label{XXZH}
\eea
where ${\cal H}_0$ is the Hamiltonian of the anisotropic spin-1/2 chain
and 
\be
{\cal H}_1= H\sum_j S^x_j. 
\label{pert}
\ee
What we do now is to bosonize at the critical point defined by ${\cal
H}_0$, i.e. the anisotropic Heisenberg chain, and then to perturb away
from this fixed point theory by (\ref{pert}). The bosonized form of
${\cal H}_0$ is
\be
{\cal H}_0 = \frac{v_s}{2}\int dx[(\partial_x\Theta)^2 +
(\partial_x\Phi)^2]\,,
\ee
where $\Phi$ is a canonical bosonic field and $\Theta$ is the dual field. 
The perturbing operator is given by
\be
{\cal H}_1=\frac{bH}{a_0}\int dx\left[\cos\left(\beta\Theta\right)
\sin\left(\frac{2\pi}{\beta}\Phi\right)\right],
\label{spin}
\ee
where
\be
\beta=\sqrt{2\pi-2{\rm arccos}(\Delta)}.
\ee
The perturbing operator \r{spin} is {\sl formally relevant}, but as
it is not a Lorentz scalar it requires special treatment. As has been
shown explicitly for the case $\Delta=0$ in Ref.[\onlinecite{shura}]
the Hamiltonian \r{XXZH} can be related to a two chain model of spinless
Luttinger liquids coupled by a weak interchain hopping. The
generalisation to $0\leq \Delta<1$ is straightforward. 
In order to match the notations of Ref.\onlinecite{shura} we perform a
shift 
\be
\Phi(x)\longrightarrow \Phi(x)+\frac{\beta}{4}\ ,
\ee
while keeping the dual field unchanged.
The 1-loop
renormalisation group analysis can then be read off from Refs
[\onlinecite{shura,yakovenko}]. At second order in the coupling the
following two scalar operators are generated radiatively
\bea
 \cos(2\beta \Theta), ~~ \cos\left(\frac{4\pi}{\beta}\Phi\right).
\eea
Altogether we thus have three perturbing operators with corresponding
dimensionless couplings $z$ (of $\cos(\beta\Theta)$
$\cos(\frac{2\pi}{\beta}\Phi)$), $g_1$ (of
$-\cos\left(\frac{4\pi}{\beta}\Phi\right)$) and $g_2$  (of
$-\cos(2\beta \Theta)$). The RG equations read 
\cite{shura,yakovenko} 
\bea
\frac{dz}{dl}&=&[2-\frac{1}{2}(\eta+\frac{1}{\eta})]z\ ,\nn
\frac{dg_1}{dl}&=&2(1-\frac{1}{\eta})g_1-(\eta-\frac{1}{\eta})z^2\
,\nn
\frac{dg_2}{dl}&=&2(1-{\eta})g_2+(\eta-\frac{1}{\eta})z^2\ ,\nn
\frac{d\ln\eta}{dl}&=&-\frac{1}{2}\left(g_2^2\eta-\frac{g_1^2}{\eta}\right)\ ,
\label{RG}
\eea
where
\be
\eta(0)=\frac{\beta^2}{2\pi}=1-\frac{1}{\pi}{\rm arccos}(\Delta).
\ee
In general these equations have to be solved numerically. However,
in the case $\eta(0)\approx 1$, i.e. a small exchange anisotropy, one can
analyze the equations by a 2-cutoff scaling procedure
\cite{shura}. This is because $z$ reaches strong coupling while
$g_{1,2}$ remain small. The gap can then be estimated as
\be
M \propto H\exp\left(-\frac{\pi}{4-4\eta(0)}\right)\ .
\ee
The gap is linear in $H$, but the coefficient is very small as
$\eta(0)\approx 1$. As is also shown in Ref.[\onlinecite{shura}], at the point
$l_0$ in the RG flow where $z$ stops remormalizing the Hamiltonian is
given by 
\bea
{\cal H}&=&\frac{v_s}{2}\int dx[(\partial_x\Theta)^2 +
(\partial_x\Phi)^2]\nn
&&+\frac{g\pi v_s}{(2\pi a_0)^2}\int dx\left[\cos(\sqrt{8\pi/\eta}\Phi)-
  \cos(\sqrt{8\pi\eta}\Theta)\right],\nn
\eea
where
\be
g=2\frac{1-\eta^{-2}}{1+\eta^2}<0
\ee
As $\eta<1$, the first cosine term is irrelevant and can be
dropped. The $\Theta$-field gets pinned at the value
$\frac{\pi}{2\beta}$, which implies that the staggered magnetization
in $y$-direction is nonzero
\be
\langle (-1)^n S^y_n\rangle \propto \langle\sin\beta\Theta\rangle={\rm
  const}. 
\ee
If $\Delta$ is not close to 1, the RG equations \r{RG} have to be
integrated numerically. Now several couplings grow simultaneously.
Eventually $\eta$ becomes very small, while $g_2$ becomes very large
and negative. This means that we again flow towards a Sine-Gordon
model for the dual field and by the same argument as before we find
that the staggered magnetization in $y$-direction is nonzero.
\section{Discussion and Conclusions}
\label{sec:disc}
As has been discussed recently in Ref.[\onlinecite{KenzelmannPRB65}],
${\rm Cs_2CoCl_4}$ is a quasi one-dimensional spin-3/2 antiferromagnet
with a strong single-ion anisotropy $D$. At energies small compared to $D$
the spin degrees of freedom are described by the anisotropic spin-1/2
Heisenberg model \r{Hamil} \cite{KenzelmannPRB65}. Due to the
smallness of the exchange constant ($J={\rm 0.23 meV}$) it is possible
to perform inelastic Neutron scattering experiments in magnetic fields
that are fairly large compared to $J$.  Hence it should be possible to
explore much of the magnetic phase diagram experimentally. 

Our analysis suggests that al low fields $H<H_c$, all components of
the dynamical structure factor are incoherent and the leading
contributions come from intermediate states with two particles. These
particles are different from the spinons of the critical XZX chain in
that they are gapped and do not carry any definite spin ($S^z$ is not
a good quantum number). In a simple picture these particles have an
interpretation as domain walls between the two possible
antiferromagnetc spin alignments in the $x$-direction. An argument in
favor of this qualitative picture may be obtained by considering the
classical line. Here it has been shown in Refs
[\onlinecite{KurmannPA112,shrock}] that the ground state is two-fold
degenerate. The expectation values of the staggered magnetization in 
the two ground states differ by an overall minus sign. This suggests
that low-lying excitations can be loosely thought of as domain walls
between the two possible ground states. 

For high fields, the $zz$-component of the dynamical structure factor
remains incoherent, whereas the $xx$, $xy$ and $yy$ components now
feature a single-particle coherent mode. For very large fields and
small $\Delta$ this particle is similar in nature to the $Z_2$ kinks
of the transverse field Ising model in the strong field limit. A
physical picture for these excitations is presented in
Ref.[\onlinecite{sachdev}]. 

It would be interesting to compare our results for the dynamical
structure factor to inelastic Neutron scattering experiments on
${\rm Cs_2CoCl_4}$ or similar compounds.

An interesting question which we have not addressed is what happens
when the magnetic field is applied at an arbitrary angle to the
exchange anisotropy. The most general case is given by
\begin{eqnarray}
{\cal H} = J \sum_j {\bf S}_j\cdot {\bf S}_{j+1}
+ (\Delta-1) S^y_j S^y_{j+1} +{\bf H}\cdot{\bf S}_j,
\end{eqnarray}
where we may set $H^x=0$ with out loss of generality (this can always
be achieved by an appropriate rotation of the quantization axis around
the $y$-direction). Rotating the spin quantization axis onto the
direction of the field leads to the following Hamiltonian
\begin{eqnarray}
{\cal H} = J \sum_j {\bf \tilde{S}}_j\cdot {\bf \tilde{S}}_{j+1}
+H\tilde{S}^z_j+ (\Delta-1){\cal H}^\prime_{j,j+1}\ ,
\end{eqnarray}
where $H={\rm sgn(H^z)}|{\bf H}|$ and
\bea
{\cal H}^\prime_{j,j+1}&=&\cos^2\theta\ \tilde{S}^y_j\tilde{S}^y_{j+1}
+\sin^2\theta\ \tilde{S}^z_j\tilde{S}^z_{j+1}\nn
&+&\sin\theta\cos\theta\ [
\tilde{S}^y_j\tilde{S}^z_{j+1}
+\tilde{S}^z_j\tilde{S}^y_{j+1}].
\label{Hprime}
\eea
Due to the presence of the $S^y_jS^z_{j+1}$ terms in \r{Hprime} the
resulting Hamiltonian can no longer be Jordan-Wigner transformed into
an expression that is local in terms of Fermion operators. Hence the
extension of the mean-field approximation to the case of an arbitrary
orientation of the magnetic field is not straightforward.

\acknowledgments
We thank D.A. Tennant and especially R. Coldea for many important
discussions and suggestions and F. Capraro for providing us
with his DMRG data. F.H.L.E. acknowledges important correspondence
with A. Furusaki and T. Hikihara. Work at Brookhaven National
Laboratory was carried out under contract number DE-AC02-98 CH10886,
Division of Material Science, U.S. Department of Energy. J.-S. Caux
thanks the Institute for Strongly Correlated and Complex Systems at
BNL for hospitality and support.


\begin{thebibliography}{99}

\bibitem{qcp}
D.A. Tennant, B. Lake, S.E. Nagler and C.D. Frost, unpublished.

\bibitem{XXZ} 
M. Takahashi, Prog. Theor. Phys. {\bf 46}, 401 (1971);
G. M\"uller, H. Thomas, H. Beck and J.C. Bonner, Phys. Rev. B{\bf 24},
1429 (1981);
H.J. Schulz, Phys. Rev. B{\bf 34}, 6372 (1986);
A. Kl\"umper, Z. Phys. {\bf 91}, 507 (1993);
A.H. Bougourzi, M. Couture and M. Kacir, Phys. Rev. B{\bf 54}, {12669} (1996);
N. Kitanine, J.-M. Maillet, N. A. Slavnov and V. Terras,
Nucl. Phys. \textbf{B} 641, 487 (2002). 

\bibitem{KurmannPA112} J. Kurmann, H. Thomas and G. M\"uller, Physica
A {\bf 112}, 235 (1982).

\bibitem{KenzelmannPRB65} M. Kenzelmann, R. Coldea, D.A. Tennant,
D. Visser, M. Hofmann, P. Smeibidl and Z. Tylczynski, Phys. Rev. B {\bf
65}, 144432 (2002).

\bibitem{dmitriev}
D.V. Dmitriev, V.Y. Krivnov and A.A. Ovchinnikov,
Phys. Rev. B{\bf 65}, 172409 (2002);
D.V.Dmitriev, V.Ya.Krivnov, A.A.Ovchinnikov, A.Langari,
JETP 95, 538 (2002).

\bibitem{NiemeijerPhysica36} Th. Niemeijer, Physica {\bf 36}, 377
(1967);  {\bf 39}, 313 (1968).

\bibitem{McCoyPRA4} B. M. McCoy, E. Barouch and D. B. Abraham,
Phys. Rev. A {\bf 4}, 2331 (1971).

\bibitem{JohnsonPRA4} J. D. Johnson and B. M. McCoy, Phys. Rev. A {\bf
4}, 2314 (1971).

\bibitem{tracy}
H.G. Vaidya and C.A. Tracy, Physica 92A, 1 (1978).

\bibitem{TaylorPRB28} J. H. Taylor and G. M\"uller, Phys. Rev. B {\bf
28}, 1529 (1983).

\bibitem{fab}
F.H.L. Essler, Phys. Rev. B{\bf 59}, 14376 (1999).

\bibitem{BarouchPRA3} E. Barouch and B. M. McCoy, Phys. Rev. A {\bf 3},
786 (1971).

\bibitem{capraro}
 F. Capraro and C. Gros, preprint cond-mat/0207279.

\bibitem{Fabr91}
K. Fabrizius, U. L\"ow and K.H. M\"utter, Phys. Rev. B{\bf 44}, 9981 (1991).

\bibitem{shrock}
G. M\"uller and R.E. Shrock, Phys. Rev. B{\bf32}, 5854 (1985).

\bibitem{esr}
M. Oshikawa and I. Affleck, Phys. Rev. B{\bf 65}, 134410 (2002).

\bibitem{lukyanov}
S. Lukyanov, Nucl. Phys. {\bf B522}, 533 (1998);
S. Lukyanov, Phys. Rev. B{\bf 59}, 11163 (1999).

\bibitem{FH}
T. Hikihara and A. Furusaki, Phys. Rev. B{\bf 63}, 134438 (2001);

\bibitem{HF2}
T. Hikihara and A. Furusaki, unpublished.

\bibitem{ET}
F.H.L. Essler and A.M. Tsvelik, Phys. Rev. B{\bf 57}, 10592 (1998).

\bibitem{vladb}
V.~E. Korepin, A.~G. Izergin, and N.~M. Bogoliubov, {\em {Quantum Inverse
  Scattering Method, Correlation Functions and Algebraic Bethe Ansatz}}
  (Cambridge University Press, 1993).

\bibitem{oa2}
I. Affleck and M. Oshikawa, Phys. Rev. B{\bf 60}, 1038 (1999).

\bibitem{LZ}
S. Lukyanov and A. Zamolodchikov, Nucl. Phys. {\bf B493}, 571 (1997).

\bibitem{destri}
C. Destri and H. de Vega, Nucl. Phys. {\bf B358}, 251 (1991).

\bibitem{smirnov}
F. Smirnov, {\it Form Factors in Completely Integrable Models of
Quantum Field Theory}, World Scientific,\\ Singapore (1992).

\bibitem{lukyanov1}
S. Lukyanov, Mod. Phys. Lett. {\bf A12}, 2911 (1997).

\bibitem{lukyanov2}
S. Lukyanov and A. Zamolodchikov, Nucl. Phys. {\bf B607}, 437 (2001).

\bibitem{shura}
A.A. Nersesyan, A. Luther and F.V. Kusmartsev, Phys. Lett. {\bf A176},
363 (1993);
A.O. Gogolin, A.A. Nersesyan and A.M. Tsvelik,{$\ \ \ $}
 {\sl Bosonization and strongly correlated systems}, Cambridge
University Press, (1998), chapter 20;

\bibitem{yakovenko}
V.M. Yakovenko, JETP Lett. {\bf 56}, 5101 (1992);

\bibitem{sachdev}
S. Sachdev, ``Quantum Phase Transitions'', Cambridge University Press,
Cambridge (1999).
\end{thebibliography}
\end{document}